%% file: paper.tex
\pdfoutput=1
\documentclass{comjnlarxiv}

\usepackage{array}
\usepackage{graphicx}
\usepackage{multirow}
\usepackage{textgreek}
\usepackage{url}

\usepackage[utf8]{inputenc}

\usepackage{amsmath}
\usepackage{amssymb}

\newcommand{\set}[1]{\ensuremath{\{ #1 \}}}
\newcommand{\abs}[1]{\ensuremath{\lvert #1 \rvert}}
\renewcommand{\complement}[1]{\ensuremath{\overline{ #1 }}}

\newcommand{\ST}{\textsf{ST}}
\newcommand{\CST}{\textsf{CST}}
\newcommand{\CSTsada}{\textsf{CST\nobreakdash-Sada}}
\newcommand{\GCT}{\textsf{GCT}}
\newcommand{\FCST}{\textsf{FCST}}
\newcommand{\CSTnpr}{\textsf{CST\nobreakdash-NPR}}
\newcommand{\RCST}{\textsf{RST}}

\newcommand{\SA}{\textsf{SA}}
\newcommand{\ISA}{\textsf{ISA}}
\newcommand{\BWT}{\textsf{BWT}}
\newcommand{\CSA}{\textsf{CSA}}
\newcommand{\FMI}{\textsf{FMI}}
\newcommand{\SSA}{\textsf{SSA}}
\newcommand{\SSArrr}{\textsf{SSA-RRR}}
\newcommand{\CSAsada}{\textsf{CSA-Sada}}
\newcommand{\RFM}{\textsf{RFM}}
\newcommand{\mSA}{\ensuremath{\mathsf{SA}}}
\newcommand{\mISA}{\ensuremath{\mathsf{ISA}}}
\newcommand{\mBWT}{\ensuremath{\mathsf{BWT}}}
\newcommand{\mF}{\ensuremath{\mathsf{F}}}
\newcommand{\mCSA}{\ensuremath{\mathsf{CSA}}}
\newcommand{\mRFM}{\ensuremath{\mathsf{RFM}}}

\newcommand{\LCP}{\textsf{LCP}}
\newcommand{\DLCP}{\textsf{DLCP}}
\newcommand{\PLCP}{\textsf{PLCP}}
\newcommand{\RLCP}{\textsf{RLCP}}
\newcommand{\LCPbyte}{\textsf{LCP\nobreakdash-byte}}
\newcommand{\LCPdac}{\textsf{LCP\nobreakdash-dac}}
\newcommand{\mLCP}{\ensuremath{\mathsf{LCP}}}
\newcommand{\mDLCP}{\ensuremath{\mathsf{DLCP}}}
\newcommand{\mPLCP}{\ensuremath{\mathsf{PLCP}}}

\newcommand{\WT}{\textsf{WT}}

\newcommand{\C}{\textsf{C}}
\newcommand{\mC}{\ensuremath{\mathsf{C}}}
\newcommand{\RLZ}{\textsf{RLZ}}

\newcommand{\LCS}{\textsf{LCS}}
\newcommand{\mLCS}{\ensuremath{\mathsf{LCS}}}
\newcommand{\mAli}{\ensuremath{\mathsf{Align}}}
\newcommand{\mCS}{\ensuremath{\complement{\mathsf{Align}}}}
\newcommand{\mleft}{\ensuremath{\mathsf{left}}}
\newcommand{\mright}{\ensuremath{\mathsf{right}}}
\newcommand{\sdarray}{\textsf{sdarray}}
\newcommand{\slarray}{\textsf{slarray}}
\newcommand{\rselect}{\textsf{rselect}}

\newcommand{\LF}{\textsf{LF}}
\newcommand{\Psiop}{\textsf{\textPsi}}
\newcommand{\find}{\textsf{find}}
\newcommand{\locate}{\textsf{locate}}
\newcommand{\extract}{\textsf{extract}}
\newcommand{\rank}{\textsf{rank}}
\newcommand{\select}{\textsf{select}}
\newcommand{\nsv}{\textsf{nsv}}
\newcommand{\nsev}{\textsf{nsev}}
\newcommand{\psv}{\textsf{psv}}
\newcommand{\psev}{\textsf{psev}}
\newcommand{\rmq}{\textsf{rmq}}

\newcommand{\mLF}{\ensuremath{\mathsf{LF}}}
\newcommand{\mPsi}{\ensuremath{\mathsf{\Psi}}}
\newcommand{\mfind}{\ensuremath{\mathsf{find}}}
\newcommand{\mlocate}{\ensuremath{\mathsf{locate}}}
\newcommand{\mextract}{\ensuremath{\mathsf{extract}}}
\newcommand{\mrank}{\ensuremath{\mathsf{rank}}}
\newcommand{\mselect}{\ensuremath{\mathsf{select}}}
\newcommand{\mlcp}{\ensuremath{\mathsf{lcp}}}
\newcommand{\mpsv}{\ensuremath{\mathsf{psv}}}
\newcommand{\mpsev}{\ensuremath{\mathsf{psev}}}
\newcommand{\mnsv}{\ensuremath{\mathsf{nsv}}}
\newcommand{\mnsev}{\ensuremath{\mathsf{nsev}}}
\newcommand{\mrmq}{\ensuremath{\mathsf{rmq}}}
\newcommand{\Oh}{\ensuremath{\mathsf{O}}}
\newcommand{\oh}{\ensuremath{\mathsf{o}}}

\newcommand{\mRoot}{\ensuremath{\mathsf{Root}}}
\newcommand{\mLeaf}{\ensuremath{\mathsf{Leaf}}}
\newcommand{\mAncestor}{\ensuremath{\mathsf{Ancestor}}}
\newcommand{\mCount}{\ensuremath{\mathsf{Count}}}
\newcommand{\mLocate}{\ensuremath{\mathsf{Locate}}}
\newcommand{\mParent}{\ensuremath{\mathsf{Parent}}}
\newcommand{\mFChild}{\ensuremath{\mathsf{FChild}}}
\newcommand{\mNSibling}{\ensuremath{\mathsf{NSibling}}}
\newcommand{\mLCA}{\ensuremath{\mathsf{LCA}}}
\newcommand{\mSDepth}{\ensuremath{\mathsf{SDepth}}}
\newcommand{\mTDepth}{\ensuremath{\mathsf{TDepth}}}
\newcommand{\mLAQ}{\ensuremath{\mathsf{LAQ}}}
\newcommand{\mSLink}{\ensuremath{\mathsf{SLink}}}
\newcommand{\mChild}{\ensuremath{\mathsf{Child}}}
\newcommand{\mLetter}{\ensuremath{\mathsf{Letter}}}

\newcommand{\zerobit}{$0$\nobreakdash-bit}
\newcommand{\mus}{\textmu{}s}

\title[Relative Suffix Trees]{Relative Suffix Trees}

\shortauthors{A. Farruggia et al.}

\author{Andrea Farruggia}
\affiliation{Department of Computer Science, University of Pisa.}

\affiliation{CeBiB -- Center for Biotechnology and Bioengineering, Chile.}
\author{Travis Gagie$^{2,}$}
\affiliation{EIT, Diego Portales University, Chile.}

\author{Gonzalo Navarro$^{2,}$}
\affiliation{Department of Computer Science, University of Chile, Chile.}

\author{Simon J. Puglisi}
\affiliation{Department of Computer Science, University of Helsinki, Finland.}

\author{Jouni Sir\'en}
\affiliation{Wellcome Trust Sanger Institute, United Kingdom.}
\email{jouni.siren@iki.fi}

\begin{abstract}
Suffix trees are one of the most versatile data structures in stringology,
with many applications in bioinformatics. Their main drawback is their size,
which can be tens of times larger than the input sequence. Much effort has
been put into reducing the space usage, leading ultimately to compressed
suffix trees. These compressed data structures can efficiently simulate the
suffix tree, while using space proportional to a compressed representation of
the sequence.
In this work, we take a new approach to compressed suffix trees for repetitive
sequence collections, such as collections of individual genomes. We compress
the suffix trees of individual sequences relative to the suffix tree of a
reference sequence. These relative data structures provide competitive
time/space trade-offs, being almost as small as the smallest compressed suffix
trees for repetitive collections, and competitive in time with the largest and
fastest compressed suffix trees. 
\end{abstract}

\begin{document}

\maketitle

\input{intro}

\input{background}

\input{rfm}

\input{rcst}

\input{exper}

\input{concl}

\section*{Funding}

This work was supported by Basal Funds FB0001, Conicyt, Chile;
Fondecyt Grant [1-170048], Chile;
Academy of Finland grants [258308] and [250345] (CoECGR);
the Jenny and Antti Wihuri Foundation, Finland;
and the Wellcome Trust grant [098051].

\bibliographystyle{compj}
\bibliography{paper}

\end{document}

%% file: intro.tex
\section{Introduction}

The \emph{suffix tree} \cite{Weiner1973} is one of the most powerful bioinformatic tools to
answer complex queries on DNA and protein sequences \cite{Gus97,Ohl13,MBCT15}.
A serious problem that hampers its wider use on large genome sequences is its
size, which may be 10--20 bytes per character. In addition, the non-local
access patterns required by most interesting problems solved with suffix trees
complicate secondary-memory deployments. This problem has led to numerous
efforts to reduce the size of suffix trees by representing them using 
\emph{compressed data structures} \cite{Sadakane2007,Fischer2009a,Ohlebusch2009,Ohlebusch2010,Fis10,Russo2011,Gog2011a,GO13b,Abeliuk2013,Navarro2014a,Navarro2015,Ock15,BCGPR15}, 
leading to \emph{compressed suffix trees} (\CST). Currently, the smallest
\CST{} is the so-called \emph{fully-compressed suffix tree} (\FCST)
\cite{Russo2011,Navarro2014a}, which uses 5 \emph{bits} per character (bpc)
for DNA sequences, but takes milliseconds to simulate suffix
tree navigation operations. In the other extreme, Sadakane's \CST{}
\cite{Sadakane2007,Gog2011a} uses about 12~bpc and operates in
microseconds, and even nanoseconds for the simplest operations.

A space usage of 12~bpc may seem reasonable to handle, for example, one human
genome, which has about 3.1 billion bases: it can be operated within a
RAM of 4.5~GB (the representation contains the sequence as well). However,
as the price of sequencing has fallen, sequencing the genomes of a large
number of individuals has become a routine activity. The \emph{1000 Genomes
Project} \cite{1000GP2015} sequenced the genomes of several thousand humans,
while newer projects can be orders of magnitude larger. This has made the
development of techniques for storing and analyzing huge amounts of sequence
data flourish.

Just storing 1000 human genomes using a 12~bpc \CST{} requires almost 4.5~TB, which
is much more than the amount of memory available in a commodity server. Assuming that
a single server has 256 GB of memory, we would need a cluster of 18 servers to
handle such a collection of \CST{}s (compared to over 100 with classical suffix
tree implementations!). With the smaller (and much slower) \FCST, this would
drop to 7--8 servers. It is clear that further space reductions in the
representation of compressed suffix trees would lead to reductions in hardware, communication,
and energy costs when implementing complex searches over large genomic
databases.

An important characteristic of those large genome databases is that they
usually consist of the genomes of individuals of the same or closely related
species. This implies that the collections are highly \emph{repetitive}, that
is, each genome can be obtained by concatenating a relatively small number of
substrings of other genomes
and adding a few new characters. When repetitiveness is considered, much higher
compression rates can be obtained in compressed suffix trees. For example, it is possible to reduce
the space to 1--2~bpc (albeit with operation times in the milliseconds)
\cite{Abeliuk2013}, or to 2--3~bpc with operation times in the microseconds
\cite{Navarro2015}. Using 2~bpc, our 1000 genomes could be handled
with just 3 servers with 256~GB of memory. 

Compression algorithms best capture repetitiveness by using \emph{grammar-based}
compression or \emph{Lempel-Ziv} compression.\footnote{We refer to ``long-range''
repetitiveness, where similar texts may be found far away in the text
collection.} In the first case \cite{KY00,CLLPPSS05} one finds a context-free
grammar that generates (only) the text collection.
Rather than compressing the text directly,
the current \CST{}s for repetitive collections \cite{Abeliuk2013,Navarro2015}
apply grammar-based compression on the data structures that simulate the suffix tree.
Grammar-based compression yields relatively easy direct access to the compressed
sequence \cite{BLRSRW15}, which makes it attractive compared to Lempel-Ziv
compression \cite{ZL77}, despite the latter generally using less space.

Lempel-Ziv compression cuts the collection into \emph{phrases}, each of which
has already appeared earlier in the collection. To extract the content of a phrase, one may have
to recursively extract the content at that earlier position, following a 
possibly long chain of indirections.
So far, the indexes built on Lempel-Ziv compression \cite{KN13} or on
combinations of Lempel-Ziv and grammar-based compression \cite{GGKNP12,GGKNP14,GP15}
support only pattern matching, which is just one of the wide range of
functionalities offered by suffix trees. The high cost to access the data
at random positions lies at the heart of the research on indexes built on
Lempel-Ziv compression.

A simple way out of this limitation is the so-called \emph{relative Lempel-Ziv}
(\RLZ) compression \cite{Kuruppu2010}, where one of the sequences is represented
in plain form and the others can only take phrases from that \emph{reference
sequence}. This enables immediate access for the symbols inside any copied
phrase (as no transitive referencing exists) and, at least if a good reference
sequence has been found, offers compression competitive with the
classical Lempel-Ziv. In our case, taking any random genome per species as the
reference is good enough; more sophisticated techniques have been studied
\cite{KPZ11,KBSCZ12,LPMW16}. Structures for direct access \cite{DG11,Ferrada2014}
and even for pattern matching \cite{DJSS14} have been developed
on top of \RLZ.

Another approach to compressing a repetitive collection while supporting interesting queries is to build an automaton that accepts the sequences in the collection, and then index the state diagram as an directed acyclic graph (DAG); see, for example,~\cite{MaciucaEtAl16,PatenEtAl17,Siren17} for recent discussions.  The first data structure to take this approach was the Generalized Compressed Suffix Array (GCSA)~\cite{SVM14,Siren17}, which was designed for pangenomics so queries can return information about sequences not in the collection but that can be obtained from those in the collection by recombination.

The FM-index of an alignment (FMA)~\cite{NaKPLLMP16,NaEtAl17} is similar to the GCSA but indexes only the sequences in the collection: whereas the GCSA conceptually embeds the automaton in a de Bruijn graph, the FMA embeds it in a coloured de Bruijn graph~\cite{IqbalEtAl12}, preserving its specificity.  Both the GCSA and FMA are practical but neither support the full functionality of a suffix tree.  The precursor to the FMA, the suffix tree of an alignment (STA)~\cite{NPCHIMP13,NPLHLMP13}, allows certain disjunctions in the suffix tree's edge labels in order to reduce the size of the tree while maintaining its functionality. Unlike the FMA, however, the STA has not been implemented.  Both the STA and FMA divide the sequences in the collection into regions of variation and conserved regions, and depend on the conserved regions being long enough that they can be distinguished from each other and the variations.  This dependency makes these structures vulnerable to even a small change in even one sequence to an otherwise-conserved region, which could hamper their scalability.

\subsection{One general $\CST$ or many individual $\CST$s}

It is important to note that the existing techniques to reduce the space of a 
collection of suffix trees on similar texts build a structure that indexes the 
collection {\em as a whole}, which is similar to concatenating all the texts of
the collection and building a single suffix tree on the concatenation. As such,
these structures do not provide the same functionality of having an individual 
\CST{} of each sequence.

Exploiting the repetitiveness of a collection while retaining separate index
structures for each text has only been achieved for a simpler pattern-matching 
index, the {\em suffix array} \cite{Manber1993}, by means of the so-called 
relative FM-indexes~\cite{Belazzougui2014}. The suffix array is a component of 
the suffix tree.

Depending on the application, we may actually need a single $\CST$ for the 
whole collection, or one for each sequence. In bioinformatics, a single $\CST$ 
is more appropriate for search and discovery of motifs across a whole 
population, for example by looking for approximate occurrences of a certain 
sequence in the genomes of the population or by discovering significant 
sequences that appear in many individuals. Other bioinformatic problems, for 
example related to the study of diseases, inheritance patterns, or forensics, 
boil down to searching or discovering patterns in the genomes 
of individuals, by finding common approximate subsequences between two genomes,
or looking for specific motifs or discovering certain patterns in a single
genome.

An example of recent research making use of the relative storage of individual genomic datasets is how Muggli et al.~\cite{Muggli2017} (see also~\cite{Alipanahi2017,Almodaresi2017}) adapted relative FM-indexes to an FM-index variant that Bowe et al.~\cite{Bowe2012} had described for de Bruijn graphs, thus obtaining a space-efficient implementation of Iqbal et al.'s~\cite{Iqbal2012} coloured de Bruijn graphs. These overlay de Bruijn graphs for many individuals to represent genetic variation in a population.

\subsection{Our contribution}

In this paper, we develop a \CST{} for repetitive collections by augmenting 
the relative FM-index with structures based on \RLZ. This turns out to be the 
first \CST{} representation that takes
advantage of the repetitiveness of the texts in a collection while at the same
time offering an individual \CST{} for each such text. Besides retaining the 
original functionality, such an approach greatly simplifies inserting and 
deleting texts in the collection and implementing the index in distributed form.

Our compressed suffix tree, called Relative Suffix Tree (\RCST), follows a 
trend of \CST{}s
\cite{Fischer2009a,Ohlebusch2009,Fis10,Ohlebusch2010,Gog2011a,Abeliuk2013} 
that use only a suffix array and an 
array with the length of the longest common prefix between each suffix and the 
previous one in lexicographic order (called \LCP). We use the relative FM-index
as our suffix array, and compress \LCP{} using \RLZ. On top of the \RLZ{} 
phrases we build a tree of range minima that enables fast range minimum 
queries, as well as next- and previous-smaller-value queries, on \LCP{} 
\cite{Abeliuk2013}. All the \CST{} functionality is built on those queries 
\cite{Fischer2009a}. Our main algorithmic contribution is this 
\RLZ\nobreakdash-based representation of the \LCP{} array with the required 
extra functionality.

On a collection of human genomes, our \RCST{} achieves less than 3~bpc and 
operates within microseconds. This performance is comparable to that of a 
previous \CST{} \cite{Navarro2015} (as explained, however,
the \RCST{} provides a different functionality because it retains the 
individual \CST{}s).

%% file: background.tex
\section{Background}

A \emph{string} $S[1,n] = s_{1} \dotso s_{n}$ is a sequence of
\emph{characters} over an \emph{alphabet} $\Sigma = \set{1, \dotsc, \sigma}$.
For indexing purposes, we often consider \emph{text} strings $T[1,n]$ that are
terminated by an \emph{endmarker} $T[n] = \$ = 0$ not occurring elsewhere in
the text. \emph{Binary} sequences are sequences over the alphabet $\set{0,1}$.
If $B[1,n]$ is a binary sequence, its \emph{complement} is binary sequence
$\complement{B}[1,n]$, with $\complement{B}[i] = 1 - B[i]$.

For any binary sequence $B[1,n]$, we define the \emph{subsequence} $S[B]$ of
string $S[1,n]$ as the concatenation of the characters $s_{i}$ with $B[i] = 1$.
The complement $\complement{S}[B]$ of subsequence $S[B]$ is the subsequence
$S[\complement{B}]$. Contiguous subsequences $S[i,j]$ are called
\emph{substrings}. Substrings of the form $S[1,j]$ and $S[i,n]$, $i,j \in
[1,n]$, are called \emph{prefixes} and \emph{suffixes}, respectively. We
define the \emph{lexicographic order} among strings in the usual way.

\subsection{Full-text indexes}

The \emph{suffix tree} (\ST)~\cite{Weiner1973} of text $T$ is a trie
containing the suffixes of $T$, with unary paths compacted into single edges.
Because the degree of every internal node is at least two, there can be at most
$2n-1$ nodes, and the suffix tree can be stored in $\Oh(n \log n)$ bits. In
practice, this is at least $10n$ bytes for small texts~\cite{Kurtz1999}, and
more for large texts as the pointers grow larger. If $v$ is a node of a suffix
tree, we write $\pi(v)$ to denote the concatenation of the labels of the path
from the root to $v$.

\emph{Suffix arrays} (\SA)~\cite{Manber1993} were introduced as a
space-efficient alternative to suffix trees. The suffix array $\mSA_{T}[1,n]$ of
text $T$ is an array of pointers to the suffixes of the text in lexicographic
order.\footnote{We drop the subscript if the text is evident from the context.}
In its basic form, the suffix array requires $n \log n$ bits in
addition to the text, but its functionality is more limited than that of the
suffix tree. In addition to the suffix array, many algorithms also use the
\emph{inverse suffix array} $\mISA[1,n]$, with $\mSA[\mISA[i]] = i$ for all
$i$.

Let $\mlcp(S_{1}, S_{2})$ be the length of the \emph{longest common prefix}
(\LCP) of strings $S_{1}$ and $S_{2}$. The \LCP{}
\emph{array}~\cite{Manber1993} $\mLCP[1,n]$ of text $T$ stores the \LCP{}
lengths for lexicographically adjacent suffixes of $T$ as $\mLCP[i] =
\mlcp(T[\mSA[i-1],n], T[\mSA[i],n])$ (with $\mLCP[1] = 0$). Let $v$ be an internal node of the
suffix tree, $\ell = \abs{\pi(v)}$ the \emph{string depth} of node $v$, and
$\mSA[sp,ep]$ the corresponding suffix array interval. The following
properties hold for the \emph{lcp-interval} $\mLCP[sp,ep]$: i) $\mLCP[sp] <
\ell$; ii) $\mLCP[i] \ge \ell$ for all $sp < i \le ep$; iii) $\mLCP[i] = \ell$
for at least one $sp < i \le ep$; and iv) $\mLCP[ep+1] <
\ell$~\cite{Abouelhoda2004}.

Abouelhoda et al.~\cite{Abouelhoda2004} showed how traversals on the suffix
tree could be simulated using the suffix array, the \LCP{} array, and a
representation of the suffix tree topology based on lcp-intervals, paving
the way for more space-efficient suffix tree representations.

\subsection{Compressed text indexes}

Data structures supporting \rank{} and \select{} queries over sequences are
the main building blocks of compressed text indexes. If $S$ is a sequence, we
define $\mrank_{c}(S,i)$ as the number of occurrences of character $c$ in
the prefix $S[1,i]$, while $\mselect_{c}(S,j)$ is the position of the occurrence
of rank $j$ in sequence $S$. A \emph{bitvector} is a representation of a
binary sequence supporting fast \rank{} and \select{} queries.
\emph{Wavelet trees} (\WT)~\cite{Grossi2003} use bitvectors to support \rank{}
and \select{} on general sequences.

The \emph{Burrows-Wheeler transform} (\BWT)~\cite{Burrows1994} is a reversible
permutation $\mBWT[1,n]$ of text $T$. It is defined as $\mBWT[i] = T[\mSA[i] -
1]$ (with $\mBWT[i] = T[n]$ if $\SA[i] = 1$). Originally intended for data
compression, the Burrows-Wheeler transform has been widely used in
space-efficient text indexes, because it shares the combinatorial structure of
the suffix tree and the suffix array.

Let \LF{} be a function such that $\mSA[\mLF(i)] = \mSA[i] - 1$ (with
$\mSA[\mLF(i)] = n$ if $\mSA[i] = 1$). We can compute it as $\mLF(i) =
\mC[\mBWT[i]] + \mrank_{\mBWT[i]}(\mBWT, i)$, where $\mC[c]$ is the number of
occurrences of characters with lexicographical values smaller than $c$ in
\BWT. The inverse function of \LF{} is $\mPsi$, with $\mPsi(i) =
\mselect_{c}(\mBWT, i - \mC[c])$, where $c$ is the largest character value
with $\mC[c] < i$. With functions $\mPsi$ and \LF, we can move forward and
backward in the text, while maintaining the lexicographic rank of the current
suffix. If the sequence $S$ is not evident from the context, we write $\mLF_{S}$
and $\mPsi_{S}$.

\emph{Compressed suffix arrays} (\CSA) \cite{Sadakane2003,Ferragina2005a,Grossi2005} are
text indexes supporting a functionality similar to the suffix array. This
includes the following queries: i) $\mfind(P) = [sp,ep]$ determines the
lexicographic range of suffixes starting with \emph{pattern} $P[1,\ell]$; ii)
$\mlocate(sp,ep) = \mSA[sp,ep]$ returns the starting positions of these
suffixes; and iii) $\mextract(i,j) = T[i,j]$ extracts substrings of the text.
In practice, the \find{} performance of {\CSA}s can be
competitive with suffix arrays, while \locate{} queries are orders of
magnitude slower~\cite{Ferragina2009a}. Typical index sizes are less than the
size of the uncompressed text.

The \emph{FM-index} (\FMI) \cite{Ferragina2005a} is a common type of
compressed suffix array. A typical implementation \cite{Ferragina2007a}
stores the \BWT{} in a
wavelet tree \cite{Grossi2003}. The index implements \find{} queries via
\emph{backward searching}. Let $[sp,ep]$ be the lexicographic
range of the suffixes of the text starting with suffix $P[i+1,\ell]$ of the
pattern. We can find the range matching suffix $P[i,\ell]$ with a
generalization of function \LF{} as
\begin{eqnarray*}
\mLF([sp,ep],P[i]) &\!\!\!=\!\!\!&
[\mC[P[i]] + \mrank_{P[i]}(\mBWT, sp\!-\!1) \!+\! 1, \\
&& \,\,\mC[P[i]] + \mrank_{P[i]}(\mBWT, ep)].
\end{eqnarray*}

We support \locate{} queries by \emph{sampling} some suffix array pointers. If
we want to determine a value $\mSA[i]$ that has not been sampled, we can
compute it as $\mSA[i] = \mSA[j]+k$, where $\mSA[j]$ is a sampled pointer
found by iterating \LF{} $k$ times, starting from position $i$. Given
\emph{sample interval} $d$, the samples can be chosen in \emph{suffix order},
sampling $\mSA[i]$ at positions divisible by $d$, or in \emph{text order},
sampling $T[i]$ at positions divisible by $d$ and marking the sampled \SA{}
positions in a bitvector. Suffix-order sampling requires less space, often
resulting in better time/space trade-offs in practice, while text-order
sampling guarantees better worst-case performance. We also sample the \ISA{}
pointers for \extract{} queries. To extract $T[i,j]$, we find the nearest
sampled pointer after $T[j]$, and traverse backwards to $T[i]$ with
function \LF.

\begin{table*}
\centering{}
\caption{Typical compressed suffix tree operations.}\label{table:cst
operations}

\begin{tabular}{ll}
\hline
\noalign{\smallskip}
\textbf{Operation}  & \textbf{Description} \\
\noalign{\smallskip}
\hline
\noalign{\smallskip}
$\mRoot()$          & The root of the tree. \\
$\mLeaf(v)$         & Is node $v$ a leaf? \\
$\mAncestor(v,w)$   & Is node $v$ an ancestor of node $w$? \\
\noalign{\smallskip}
$\mCount(v)$        & Number of leaves in the subtree with $v$ as the root. \\
$\mLocate(v)$       & Pointer to the suffix corresponding to leaf $v$. \\
\noalign{\smallskip}
$\mParent(v)$       & The parent of node $v$. \\
$\mFChild(v)$       & The first child of node $v$ in alphabetic order. \\
$\mNSibling(v)$     & The next sibling of node $v$ in alphabetic order. \\
$\mLCA(v,w)$        & The lowest common ancestor of nodes $v$ and $w$. \\
\noalign{\smallskip}
$\mSDepth(v)$       & \emph{String depth}: Length $\ell = \abs{\pi(v)}$ of the
label from the root to node $v$. \\
$\mTDepth(v)$       & \emph{Tree depth}: The depth of node $v$ in the suffix
tree. \\
$\mLAQ_{S}(v,d)$    & The highest ancestor of node $v$ with string depth at
least $d$. \\
$\mLAQ_{T}(v,d)$    & The ancestor of node $v$ with tree depth $d$. \\
\noalign{\smallskip}
$\mSLink(v)$        & \emph{Suffix link}: Node $w$ such that $\pi(v) = c \pi(w)$ for
a character $c \in \Sigma$. \\
$\mSLink^{k}(v)$    & Suffix link iterated $k$ times. \\
\noalign{\smallskip}
$\mChild(v,c)$      & The child of node $v$ with edge label starting with
character $c$. \\
$\mLetter(v,i)$     & The character $\pi(v)[i]$. \\
\noalign{\smallskip}
\hline
\end{tabular}
\end{table*}

\emph{Compressed suffix trees} (\CST) \cite{Sadakane2007} are compressed text
indexes supporting the full functionality of a suffix tree (see
Table~\ref{table:cst operations}). They combine a compressed suffix array, a
compressed representation of the \LCP{} array, and a compressed representation
of suffix tree topology. For the \LCP{} array, there are several common
representations:
\begin{itemize}
\item \LCPbyte{} \cite{Abouelhoda2004} stores the \LCP{} array as a byte
array. If $\mLCP[i] < 255$, the \LCP{} value is stored in the byte array.
Larger values are marked with a $255$ in the byte array and stored separately.
As many texts produce small \LCP{} values, \LCPbyte{} usually requires
$n$ to $1.5n$ bytes of space.
\item We can store the \LCP{} array by using variable-length codes. \LCPdac{}
uses \emph{directly addressable codes} \cite{Brisaboa2009} for the purpose,
resulting in a structure that is typically somewhat smaller and somewhat
slower than \LCPbyte.
\item The \emph{permuted} \LCP{} (\PLCP) \emph{array} \cite{Sadakane2007}
$\mPLCP[1,n]$ is the \LCP{} array stored in text order and used as $\mLCP[i] =
\mPLCP[\mSA[i]]$. Because $\mPLCP[i+1] \ge \mPLCP[i]-1$, the array can be
stored as a bitvector of length $2n$ in $2n+\oh(n)$ bits. If the text is
repetitive, run-length encoding can be used to compress the bitvector to take
even less space \cite{Fischer2009a}. Because accessing \PLCP{} uses \locate,
it is much slower than the above two encodings.
\end{itemize}

Suffix tree topology representations are the main difference between the
various \CST{} proposals. While the compressed suffix arrays and the \LCP{} arrays
are interchangeable, the tree representation determines how various suffix tree
operations are implemented. There are three main families of compressed suffix
trees:
\begin{itemize}
\item \emph{Sadakane's compressed suffix tree} (\CSTsada) \cite{Sadakane2007}
uses a \emph{balanced parentheses} representation for the tree. Each node is
encoded as an opening parenthesis, followed by the encodings of its children
and a closing parenthesis. This can be encoded as a bitvector of length $2n'$,
where $n'$ is the number of nodes, requiring up to $4n+\oh(n)$ bits.
\CSTsada{} tends to be larger and faster than the other compressed suffix
trees \cite{Gog2011a,Abeliuk2013}.
\item The \emph{fully compressed suffix tree} (\FCST) of Russo et
al.~\cite{Russo2011,Navarro2014a} aims to use as little space as possible. It
does not require an \LCP{} array at all, and stores a balanced parentheses
representation for a sampled subset of suffix tree nodes in $\oh(n)$ bits.
Unsampled nodes are retrieved by following suffix links. \FCST{} is smaller
and much slower than the other compressed suffix trees
\cite{Russo2011,Abeliuk2013}.
\item Fischer et al.~\cite{Fischer2009a} proposed an intermediate
representation, \CSTnpr, based on lcp-intervals. Tree navigation is handled
by searching for the values defining the lcp-intervals. \emph{Range minimum
queries} $\mrmq(sp,ep)$ find the leftmost minimal value in $\mLCP[sp,ep]$,
while \emph{next/previous smaller value} queries $\mnsv(i)$/$\mpsv(i)$ find
the next/previous \LCP{} value smaller than $\mLCP[i]$. After the improvements
by various authors \cite{Ohlebusch2009,Fis10,Ohlebusch2010,Gog2011a,Abeliuk2013},
the \CSTnpr{} is perhaps the most practical compressed suffix tree.
\end{itemize}

For typical texts and component choices, the size of compressed suffix trees
ranges from the $1.5n$ to $3n$ bytes of \CSTsada{} to the $0.5n$ to $n$ bytes
of \FCST{} \cite{Gog2011a,Abeliuk2013}. There are also some \CST{} variants
for repetitive texts, such as versioned document collections and collections
of individual genomes. Abeliuk et al.~\cite{Abeliuk2013} developed a variant
of \CSTnpr{} that can sometimes be smaller than $n$ bits, while achieving
performance similar to the \FCST. Navarro and Ordóñez \cite{Navarro2015} used
grammar-based compression for the tree representation of \CSTsada. The
resulting compressed suffix tree (\GCT) requires slightly more space than the
\CSTnpr{} of Abeliuk et al., while being closer to the non-repetitive
\CSTsada{} and \CSTnpr{} in performance.

\subsection{Relative Lempel-Ziv}\label{sect:rlz}

\emph{Relative Lempel-Ziv} (\RLZ) parsing \cite{Kuruppu2010} compresses
\emph{target} sequence $S$ relative to \emph{reference} sequence $R$. The
target sequence is represented as a concatenation of $z$ \emph{phrases} $w_{i}
= (p_{i}, \ell_{i}, c_{i})$, where $p_{i}$ is the starting position of the
phrase in the reference, $\ell_{i}$ is the length of the copied substring, and
$c_{i}$ is the \emph{mismatch} character. If phrase $w_{i}$ starts from
position $p'$ in the target, then $S[p',p'+\ell_{i}-1] =
R[p_{i},p_{i}+\ell_{i}-1]$ and $S[p'+\ell_{i}] = c_{i}$.

The shortest \RLZ{} parsing of the target sequence can be found in
(essentially) linear time. The algorithm builds a \CSA{} for the reverse of
the reference sequence, and then parses the target sequence greedily by using
backward searching. If the edit distance between the reference and the target
is $s$, we need at most $s$ phrases to represent the target sequence. On the
other hand, because the relative order of the phrases can be different in
sequences $R$ and $S$, the edit distance can be much larger than the number of
phrases in the shortest \RLZ{} parsing.

In a straightforward implementation, the \emph{phrase pointers} $p_{i}$ and
the mismatch characters $c_{i}$ can be stored in arrays $W_{p}$ and
$W_{c}$. These arrays take $z \log \abs{R}$ and $z \log \sigma$ bits,
respectively. To support random access to the target sequence, we can encode
phrase lengths as a bitvector $W_{\ell}$ of length $\abs{S}$ \cite{Kuruppu2010}:
we set $W_{\ell}[j] = 1$ if $S[j]$ is the first character of a phrase. The
bitvector requires $z \log \frac{n}{z} + \Oh(z)$ bits if we use the
\sdarray{} representation \cite{Okanohara2007}. To extract $S[j]$, we first
determine the phrase $w_{i}$, with $i = \mrank_{1}(W_{\ell}, j)$. If
$W_{\ell}[j+1] = 1$, we return the mismatch character $W_{c}[i]$. Otherwise
we determine the phrase offset with a \select{} query, and return the character
$R[W_{p}[i] + j - \mselect_{1}(W_{\ell}, i)]$.

Ferrada et al.~\cite{Ferrada2014} showed how, by using \emph{relative pointers} instead of absolute pointers, we can avoid the use of \select{} queries. They also achieved better compression of DNA collections, in which most of the differences between the target sequences and the reference sequence are single-character \emph{substitutions}. By setting $W_{r}[i] = p_{i} - \mselect_{1}(W_{\ell}, i)$, the general case simplifies to $S[j] = R[W_{r}[i] + j]$.  If most of the differences are single-character substitutions, $p_{i+1}$ will often be $p_{i} + \ell_{i} + 1$. This corresponds to $W_{r}[i+1] = W_{r}[i]$ with relative pointers, making \emph{run-length encoding} of the pointer array worthwhile.

When we sort the suffixes in lexicographic order, substitutions in the text move suffixes around, creating \emph{insertions} and \emph{deletions} in the suffix array and related structures. In the \LCP{} array, an insertion or deletion affecting $\mLCP[i]$ can also change the value of $\mLCP[i+1]$. Hence \RLZ{} with relative pointers is not enough to compress the \LCP{} array.

Cox et al.~\cite{CoxEtAl16} modified Ferrada et al.'s version of \RLZ{} to handle other small variations in addition to single-character substitutions. After adding a phrase to the parse, we look ahead a bounded number of positions to find potential phrases with a relative pointer $W_{r}[i]$ close to the previous \emph{explicit} relative pointer $W_{r}[j]$. If we can find a sufficiently long phrase this way, we encode the pointer \emph{differentially} as $W_{r}[i] - W_{r}[j]$. Otherwise we store $W_{r}[i]$ explicitly. We can then save space by storing the differential pointers separately using less bits per pointer. Because there can be multiple mismatch characters between phrases $i$ and $i+1$, we also need a prefix-sum data structure $L$ for finding the range $W_{c}[a,b]$ containing the mismatches. Cox et al.\ showed that their approach compresses both DNA sequences and \LCP{} arrays better than Ferrada et al.'s version, albeit with slightly slower random access. We refer the reader to their paper for more details of their implementation.

%% file: rfm.tex
\section{Relative FM-index}
\label{sec:rfm}

The \emph{relative FM-index} (\RFM) \cite{Belazzougui2014} is a compressed
suffix array of a sequence relative to the \CSA{} of another sequence.
The index is based on approximating the
\emph{longest common subsequence} (\LCS) of $\mBWT_{R}$ and $\mBWT_{S}$,
where $R$ is the reference sequence and $S$ is the target sequence, and
storing several structures based on the common subsequence. Given a
representation of $\mBWT_{R}$ supporting \rank{} and \select{}, we can use the
relative index $\mRFM_{S \mid R}$ to simulate \rank{} and \select{} on
$\mBWT_{S}$.

In this section, we describe the relative FM-index using the notation and the
terminology of this paper. We also give an explicit description of the
\locate{} and \extract{} functionality, which was not included in the original
paper. Finally, we describe a more space-efficient variant of the algorithm
for building a relative FM-index with full functionality.

\subsection{Basic index}

Assume that we have found a long common subsequence of sequences $X$ and $Y$.
We call positions $X[i]$ and $Y[j]$ \emph{lcs-positions}, if they are in the
common subsequence. If $B_{X}$ and $B_{Y}$ are the binary sequences marking
the common subsequence ($X[\select_1(B_{X},i)] = Y[\select_1(B_{Y},i)]$), we
can move between lcs-positions in the two sequences with \rank{} and \select{}
operations. If $X[i]$ is an lcs-position, the corresponding position in
sequence $Y$ is $Y[\mselect_{1}(B_{Y}, \mrank_{1}(B_{X}, i))]$. We denote this
pair of \emph{lcs-bitvectors} $\mAli(X,Y) = \langle B_X,B_Y \rangle$.

In its most basic form, the relative FM-index $\mRFM_{S \mid R}$ only supports
\find{} queries by simulating \rank{} queries on $\mBWT_{S}$. It does this by
storing $\mAli(\BWT_{R},\BWT_{S})$ and the complements (subsequences of
non-aligned characters) $\mCS(\mBWT_{R})$ and $\mCS(\mBWT_{S})$. The
lcs-bitvectors are compressed using \emph{entropy-based compression}
\cite{Raman2007}, while the complements are stored in structures similar to
the reference $\mBWT_{R}$.

To compute $\mrank_{c}(\mBWT_{S}, i)$, we first determine the number of
lcs-positions in $\mBWT_{S}$ up to position $S[i]$ with $k =
\mrank_{1}(B_{\mBWT_{S}}, i)$. Then we find the lcs-position $k$ in $\mBWT_{R}$
with $j = \mselect_{1}(B_{\mBWT_{R}}, k)$. With these positions, we can compute
\begin{eqnarray*}
\mrank_{c}(\mBWT_{S}, i) &=& \mrank_{c}(\mBWT_{R}, j) \\
   && - \mrank_{c}(\mCS(\mBWT_{R}),j-k) \\
   && + \mrank_{c}(\mCS(\mBWT_{S}), i-k).
\end{eqnarray*}

\subsection{Relative select}

We can implement the entire functionality of a compressed suffix array with
\rank{} queries on the \BWT. However, if we use the \CSA{} in a compressed
suffix tree, we also need \select{} queries to support \emph{forward
searching} with $\mPsi$ and $\mChild$ queries. We can always implement
\select{} queries by binary searching with \rank{} queries, but the result
will be much slower than the \rank{} queries.

A faster alternative to support \select{} queries in the relative FM-index
is to build a \emph{relative select} structure \rselect{}~\cite{Boucher2015}.
Let $\mF_{X}$ be a sequence consisting of the characters of sequence $X$ in
sorted order. Alternatively, $\mF_{X}$ is a sequence such that $\mF_{X}[i] =
\mBWT_{X}[\mPsi_X(i)]$. The relative select structure consists of bitvectors
$\mAli(\mF_{R}, \mF_{S})$, where $B_{\mF_{R}}[i] = B_{\mBWT_{R}}[\mPsi_R(i)]$ 
and $B_{\mF_{S}}[i] = B_{\mBWT_{S}}[\mPsi_S(i)]$, as well as the \C{} array
$\mC_{\mLCS}$ for the common subsequence.

To compute $\mselect_{c}(\mBWT_{S}, i)$, we first determine how many of
the first $i$ occurrences of character $c$ are lcs-positions with $k =
\mrank_{1}(B_{\mF_{S}}, \mC_{\mBWT_{S}}[c] + i) - \mC_{\mLCS}[c]$. Then we check
from bit $B_{\mF_{S}}[\mC_{\mBWT_{S}}[c] + i]$ whether the occurrence we are
looking for is an lcs-position or not. If it is,
we find the position in $\mBWT_{R}$ as $j = \mselect_{c}(\mBWT_{R},
\mselect_{1}(B_{\mF_{R}}, \mC_{\mLCS}[c] + k)- \mC_{R}[c])$, and then map $j$ to
$\mselect_{c}(\mBWT_{S}, i)$ by using $\mAli(\mBWT_{R}, \mBWT_{S})$. Otherwise we
find the occurrence in $\mCS(\mBWT_{S})$ with $j = \mselect_{c}(\mCS(\mBWT_{S}),
i-k)$, and return $\mselect_{c}(\mBWT_{S}, i) = \mselect_{0}(B_{\mBWT_{S}}, j)$.

\subsection{Full functionality}

If we want the relative FM-index to support \locate{} and \extract{} queries,
we cannot build it from any common subsequence of $\mBWT_{R}$ and $\mBWT_{S}$.
We need a \emph{bwt-invariant subsequence} \cite{Belazzougui2014}, where the
alignment of the \BWT{}s is also an alignment of the original sequences.

\begin{definition}\label{def:bwt-invariant}
Let $X$ be a common subsequence of $\mBWT_{R}$ and $\mBWT_{S}$, and let
$\mBWT_{R}[i_{R}]$ and $\mBWT_{S}[i_{S}]$ be the lcs-positions corresponding to
$X[i]$. Subsequence X is bwt-invariant if
$$
\mSA_{R}[i_{R}] < \mSA_{R}[j_{R}] \iff \mSA_{S}[i_{S}] < \mSA_{S}[j_{S}]
$$
for all positions $i, j \in \set{1, \dotsc, \abs{X}}$.
\end{definition}

In addition to the structures already mentioned, the full relative FM-index
has another pair of lcs-bitvectors, $\mAli(R,S)$, which marks the
bwt-invariant subsequence in the original sequences. If $\mBWT_{R}[i_{R}]$ and
$\mBWT_{S}[i_{S}]$ are lcs-positions, we set $B_{R}[\mSA_{R}[i_{R}]-1] = 1$ and
$B_{S}[\mSA_{S}[i_{S}]-1] = 1$.\footnote{For simplicity, we assume that the
endmarker is not a part of the bwt-invariant subsequence. Hence $\mSA[i] > 1$
for all lcs-positions $\mBWT[i]$.}

To compute the answer to a $\mlocate(i)$ query, we start by iterating
$\mBWT_{S}$ backwards with \LF{} queries, until we find an lcs-position
$\mBWT_{S}[i']$ after $k$ steps. Then we map position $i'$ to the corresponding
position $j'$ in $\mBWT_{R}$ by using $\mAli(\mBWT_{R},\mBWT_{S})$. Finally we
determine $\mSA_{R}[j']$ with a \locate{} query in the reference index, and map
the result to $\mSA_{S}[i']$ by using $\mAli(R,S)$.\footnote{If $\mBWT_{S}[i']$
and $\mBWT_{R}[j']$ are lcs-positions, the corresponding lcs-positions in the
original sequences are $S[\mSA_{S}[i']-1]$ and $R[\mSA_{R}[j']-1]$.} The result
of the $\mlocate(i)$ query is $\mSA_{S}[i']+k$.

The $\mISA_{S}[i]$ access required for \extract{} queries is supported in a
similar way. We find the lcs-position $S[i+k]$ for the smallest $k \ge 0$, and
map it to the corresponding position $R[j]$ by using $\mAli(R,S)$. Then we
determine $\mISA_{R}[j+1]$ by using the reference index, and map it back to
$\mISA_{S}[i+k+1]$ with $\mAli(\mBWT_{R},\mBWT_{S})$. Finally we iterate
$\mBWT_{S}$ $k+1$ steps backward with \LF{} queries to find $\mISA_{S}[i]$.

If the target sequence contains long
insertions not present in the reference, we may also want to include
some \SA{} and \ISA{} samples for querying those regions.

\subsection{Finding a bwt-invariant subsequence}

With the basic relative FM-index, we approximate the longest common
subsequence of $\mBWT_{R}$ and $\mBWT_{S}$ by partitioning the \BWT{}s according
to lexicographic contexts, finding the longest common subsequence for each
pair of substrings in the partitioning, and concatenating the results. The
algorithm is fast, easy to parallelize, and quite space-efficient. As such,
\RFM{} construction is practical, having been tested with datasets of hundreds
of gigabytes in size.

In the following, we describe a more space-efficient variant of the original
algorithm \cite{Belazzougui2014} for finding a bwt-invariant subsequence. We
\begin{itemize}
\item save space by simulating the \emph{mutual suffix array} $\mSA_{RS}$ with
$\mCSA_{R}$ and $\mCSA_{S}$;
\item \emph{match} suffixes of $R$ and $S$ only if they are adjacent in
$\mSA_{RS}$; and
\item run-length encode the match arrays to save space.
\end{itemize}

\begin{definition}
Let $R$ and $S$ be two sequences, and let $\mSA = \mSA_{RS}$ and $\mISA =
\mISA_{RS}$. The \emph{left match} of suffix $R[i,\abs{R}]$ is the suffix
$S[\mSA[\mISA[i]-1] - \abs{R}, \abs{S}]$, if $\mISA[i] > 1$ and
$\mSA[\mISA[i]-1]$ points to a suffix of $S$ ($\mSA[\mISA[i]-1] > \abs{R}$).
The \emph{right match} of suffix $R[i,\abs{R}]$ is the suffix
$S[\mSA[\mISA[i]+1] - \abs{R}, \abs{S}]$, if $\mISA[i] < \abs{RS}$ and
$\mSA[\mISA[i]+1]$ points to a suffix of $S$.
\end{definition}

We simulate the mutual suffix array $\mSA_{RS}$ with $\mCSA_{R}$, $\mCSA_{S}$,
and the \emph{merging bitvector} $B_{R,S}$ of length $\abs{RS}$. We set
$B_{R,S}[i] = 1$, if $\mSA_{RS}[i]$ points to a suffix of $S$. The merging
bitvector can be built in $\Oh(\abs{S} \cdot t_{\mLF})$ time, where $t_{\mLF}$ is
the time required for an \LF{} query, by extracting $S$ from $\mCSA_{S}$ and
backward searching for it in $\mCSA_{R}$ \cite{Siren2009}. Suffix
$R[i,\abs{R}]$ has a left (right) match, if $B_{R,S}[\mselect_{0}(B_{R,S},
\mISA_{R}[i])-1] = 1$ ($B_{R,S}[\mselect_{0}(B_{R,S}, \mISA_{R}[i])+1] = 1)$).

Our next step is building the \emph{match arrays} $\mleft$ and $\mright$,
which correspond to the arrays $A[\cdot][2]$ and $A[\cdot][1]$ in the original
algorithm. This is done by traversing $\mCSA_{R}$ backwards from
$\mISA_{R}[\abs{R}] = 1$ with \LF{} queries and following the left and the
right matches of the current suffix. During the traversal, we maintain
the invariant $j = \mSA_{R}[i]$ with $(i,j) \leftarrow (\mLF_{R}(i), j-1)$. If
suffix $R[j,\abs{R}]$ has a left (right) match, we use the shorthand $l(j) =
\mrank_{1}(B_{R,S}, \mselect_{0}(B_{R,S}, i)-1)$ ($r(j) = \mrank_{1}(B_{R,S},
\mselect_{0}(B_{R,S}, i)+1)$) to refer to its position in $\mCSA_{S}$.

We say that suffixes $R[j,\abs{R}]$ and $R[j+1,\abs{R}]$ have the same left
match if $l(j) = \mLF_{S}(l(j+1))$. Let $R[j,\abs{R}]$ to $R[j+\ell,\abs{R}]$
be a maximal run of suffixes having the same left match, with suffixes
$R[j,\abs{R}]$ to $R[j+\ell-1,\abs{R}]$ starting with the same characters as
their left matches.\footnote{The first character of a suffix can be determined
by using the $\mC$ array.} We find the left match of suffix $R[j,\abs{R}]$ as
$j' = \mSA_{S}[l(j)]$ by using $\mCSA_{S}$, and set $\mleft[j,j+\ell-1] =
[j',j'+\ell-1]$. The right match array $\mright$ is built in a similar way.

The match arrays require $2\abs{R} \log \abs{S}$ bits of space. If sequences
$R$ and $S$ are similar, the runs in the arrays tend to be long. Hence we can
run-length encode the match arrays to save space. The traversal takes
$\Oh(\abs{R} \cdot (t_{\mLF} + t_{\mrank} + t_{\mselect}) + rd \cdot
t_{\mLF})$ time, where $t_{\mrank}$ and $t_{\mselect}$ denote the time
required by \rank{} and \select{} operations, $r$ is the number of runs in the
two arrays, and $d$ is the suffix array sample interval in
$\mCSA_{S}$.\footnote{The time bound assumes text-order sampling.}

The final step is determining the bwt-invariant subsequence. We find a
binary sequence $B_{R}[1,\abs{R}]$, which marks the common subsequence in $R$,
and a strictly increasing integer sequence $Y$, which contains the positions
of the common subsequence in $S$. This can be done by finding the longest
increasing subsequence over $R$, where we consider both $\mleft[i]$ and
$\mright[i]$ as candidates for the value at position $i$, and using the found
subsequence as $Y$. If $Y[j]$ comes from $\mleft[i]$ ($\mright[i]$), we set
$B_{R}[i] = 1$, and align suffix $R[i, \abs{R}]$ with its
left (right) match $S[Y[j], \abs{S}]$ in the bwt-invariant subsequence. We can
find $B_{R}$ and $Y$ in $\Oh(\abs{R} \log \abs{R})$ time with
$\Oh(\abs{R} \log \abs{R})$ bits of additional working space with a
straightforward modification of the dynamic programming algorithm for finding
the longest increasing subsequence. The dynamic programming tables can be
run-length encoded, but we found that this did not yield good time/space
trade-offs.

As sequence $Y$ is strictly increasing, we can convert it into binary sequence
$B_{S}[1,\abs{S}]$, marking $B_{S}[Y[j]] = 1$ for all $j$.
Afterwards, we consider the binary sequences $B_{R}$ and $B_{S}$ as the
lcs-bitvectors $\mAli(R,S)$. Because every suffix of $R$ starts with the same
character as its matches stored in the $\mleft$ and $\mright$ arrays,
subsequences $R[B_{R}]$ and $S[B_{S}]$ are identical.

For any $i$, let $i_{R} = \mselect_{1}(B_{R}, i)$ and $i_{S} =
\mselect_{1}(B_{S}, i)$ be the lcs-positions of rank $i$. As suffixes
$R[i_{R}, \abs{R}]$ and $S[i_{S}, \abs{S}]$ are aligned in the bwt-invariant
subsequence, they are also adjacent in the mutual suffix array $\mSA_{RS}$.
Hence
$$
\mISA_{R}[i_{R}] < \mISA_{R}[j_{R}] \iff \mISA_{S}[i_{S}] < \mISA_{S}[j_{S}]
$$
for $1 \le i,j \le \abs{Y}$, which is equivalent to the condition in
Definition~\ref{def:bwt-invariant}. We can convert $\mAli(R,S)$ to
$\mAli(\mBWT_{R},\mBWT_{S})$ in $\Oh((\abs{R}+\abs{S}) \cdot t_{\mLF})$ time by
traversing $\mCSA_{R}$ and $\mCSA_{S}$ backwards. The resulting subsequence of
$\mBWT_{R}$ and $\mBWT_{S}$ is bwt-invariant.

Note that the full relative FM-index is more limited than the basic index,
because it does not handle \emph{substring moves} very well. Let $R = xy$ and
$S = yx$, for two random sequences $x$ and $y$ of length $n/2$ each. Because
$\mBWT_{R}$ and $\mBWT_{S}$ are very similar, we can expect to find a common
subsequence of length almost $n$. On the other hand, the length of the longest
bwt-invariant subsequence is around $n/2$, because we can either match the
suffixes of $x$ or the suffixes of $y$ in $R$ and $S$, but not both.

%% file: rcst.tex
\section{Relative suffix tree}

The \emph{relative suffix tree} (\RCST) is a \CSTnpr{} of the
target sequence relative to a \CST{} of the reference sequence. It consists of
two major components: the relative FM-index with full functionality and the
\emph{relative} \LCP{} (\RLCP) \emph{array}. The optional relative select
structure can be generated or loaded from disk to speed up algorithms based on
forward searching. The \RLCP{} array is based on \RLZ{} parsing, while the
support for \nsv/\psv/\rmq{} queries is based on a minima tree over the
phrases.

\subsection{Relative \LCP{} array}

Given \LCP{} array $\mLCP[1,n]$, we define the \emph{differential} \LCP{}
\emph{array} $\mDLCP[1,n]$ as $\mDLCP[1] = \mLCP[1]$ and $\mDLCP[i] = \mLCP[i]
- \mLCP[i-1]$ for $i > 1$. If $\mBWT[i,j] = c^{j+1-i}$ for some $c \in
\Sigma$, then $\mLCP[\mLF(i)+1,\mLF(j)]$ is the same as $\mLCP[i+1,j]$, with
each value incremented by $1$ \cite{Fischer2009a}. This means
$\mDLCP[\mLF(i)+2,\mLF(j)] = \mDLCP[i+2,j]$, making the \DLCP{} array of a
repetitive text compressible with grammar-based compression
\cite{Abeliuk2013}.

We make a similar observation in the relative setting. If target sequence $S$
is similar to the reference sequence $R$, then their \LCP{} arrays should also
be similar. If there are long identical ranges $\mLCP_{R}[i,i+k] =
\mLCP_{S}[j,j+k]$, the corresponding \DLCP{} ranges $\mDLCP_{R}[i+1,i+k]$ and
$\mDLCP_{S}[j+1,j+k]$ are also identical. Hence we can use \RLZ{} parsing to
compress either the original \LCP{} array or the \DLCP{} array.

While the identical ranges are a bit longer in the \LCP{} array, we opt to
compress the \DLCP{} array, because it behaves better when there are long
repetitions in the sequences. In particular, assembled genomes often have long
runs of character $N$, which correspond to regions of very large \LCP{}
values. If the runs are longer in the target sequence than in the reference
sequence, the \RLZ{} parsing of the \LCP{} array will have many
mismatch characters. The corresponding ranges in the
\DLCP{} array typically consist of values $\set{-1, 0, 1}$, making them much
easier to compress.

We consider \DLCP{} arrays as strings over an integer alphabet and create an \RLZ{} parsing
of $\mDLCP_{S}$ relative to $\mDLCP_{R}$. After parsing, we switch to using
$\mLCP_{R}$ as the reference. The reference is stored in a
structure we call \slarray, which is a variant of \LCPbyte.
\cite{Abouelhoda2004}. Small values $\mLCP_{R}[i] < 255$ are stored in a byte
array, while large values $\mLCP_{R}[i] \ge 255$ are marked with a $255$ in the
byte array and stored separately. To quickly find the large values, we also
build a $\mrank_{255}$ structure over the byte array. The \slarray{} provides
reasonably fast random access and fast sequential access to the
underlying array.

The \RLZ{} parsing produces a sequence of phrases $w_{i} = (p_{i}, \ell_{i},
c_{i})$ (see Section~\ref{sect:rlz}; since we are using Cox et al.'s version, $c_{i}$ is now a string).
Because some queries involve decompressing an entire phrase, we limit the maximum phrase length to $1024$.
We also require that $\abs{c_{i}} > 0$ for all $i$, using the last character of the copied substring
as a mismatch if necessary.

Phrase lengths are encoded in the $W_{\ell}$ bitvector in the usual way. We convert the strings of mismatching
\DLCP{} values $c_{i}$ into strings of absolute \LCP{} values, append them into the mismatch array $W_{c}$, and
store the array as an \slarray. The mismatch values are used as \emph{absolute
samples} for the differential encoding.

To access $\mLCP_{S}[j]$, we determine the phrase $w_{i}$ as usual, and check whether we should return a mismatch character. If so, we compute which one using a prefix sum query on $L$, and return it.  If not, we determine the starting positions $p_{i}$ and $s_{i}$ of the phrase $w_{i}$ in the reference and the target, respectively. We can then compute the solution as
\begin{align*}
\mLCP_{S}[j] &= \mLCP_{S}[s_{i} - 1] + \sum_{k = s_{i}}^{j} \mDLCP_{S}[k] \\
&= \mLCP_{S}[s_{i} - 1] + \sum_{k = p_{i}}^{j'} \mDLCP_{R}[k] \\
&= \mLCP_{S}[s_{i} - 1] + \mLCP_{R}[j'] - \mLCP_{R}[p_{i} - 1],
\end{align*}
where $j' = p_{i} + j - s_{i}$.
Each \RLZ{} phrase ends with at least one mismatch character, so $\mLCP_{S}[s_{i} - 1]$ is readily available. After finding $\mLCP_{S}[j]$, accessing $\mLCP_{S}[j-1]$ and $\mLCP_{S}[j+1]$ is fast, as long as we do not cross phrase boundaries.

\begin{figure}
\centerline{\includegraphics[width=0.4\textwidth]{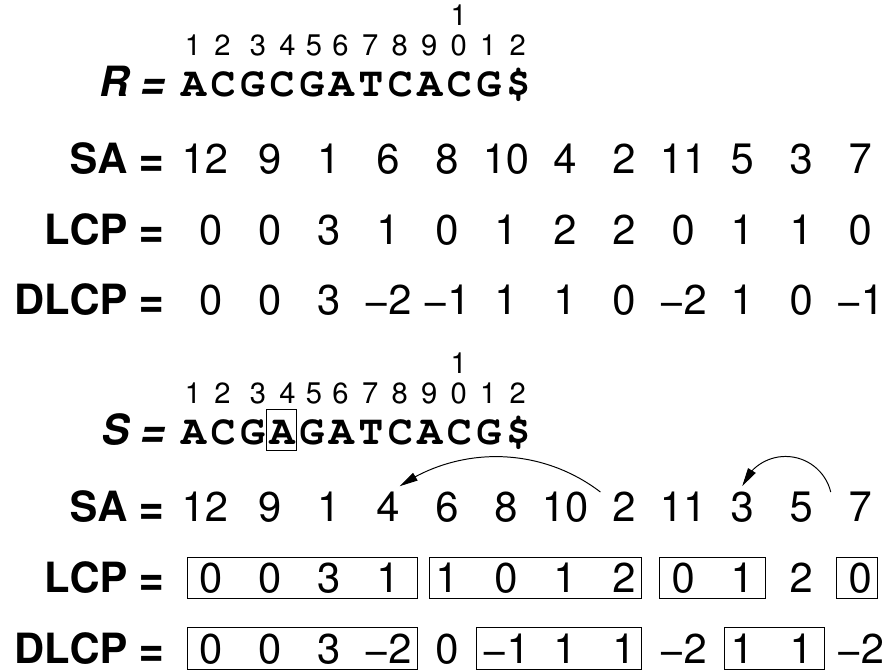}}
\caption{An example of our $\RLZ$ compression of $\DLCP$.}
\label{fig:ex}
\end{figure}

\paragraph{Example.}
Figure~\ref{fig:ex} shows an example reference sequence $R$ and target
sequence $S$, with their corresponding arrays $\SA$, $\LCP$, and $\DLCP$.
The single edit at $S[4]$ with respect to $R[4]$ may affect the positions
of suffixes $4$ and previous ones in $\SA$, although in general only a limited
number of preceding suffixes are affected. In our example, suffix $4$ moves
from position $7$ in $\SA_R$ to position $4$ in $\SA_S$, and suffix $3$ moves
from position $11$ in $\SA_R$ to position $10$ in $\SA_S$. Each suffix that
is moved from $\SA_R[i]$ to $\SA_S[j]$ may alter the values at positions $i$ 
or $i+1$ (depending on whether $j>i$ or $j<i$), as well as $j$ and $j+1$, of
$\LCP_S$. We have surrounded in rectangles the conserved regions in $\LCP_S$
(some are conserved by chance). Even some suffixes that 
are not moved may change their $\LCP$ values. In turn, each change in 
$\LCP_S[k]$ may change values $\DLCP_S[k]$ and $\DLCP_S[k+1]$.

After the change, we can parse $\DLCP_S$ into three phrases (with the copied
symbols surrounded by rectangles):
$(1,4,0)$, $(5,3,-2)$, $(6,2,-2)$, where the latter is formed by chance.
We represent this parsing as $W_c = \langle 1,0,0 \rangle$ (since we store the
absolute $\LCP_S$ values for the mismatches), $W_\ell = 100001000100$, and
$W_p = \langle 1,5,6 \rangle$ (or rather $W_r = \langle 0,-1,-4 \rangle$).

Let us compute $\LCP_S[j]$ for $j=8$. This corresponds to phrase number
$i = \rank(W_\ell,j) = 2$, which starts at position $s_i = \select(W_\ell,i)=6$
in $\LCP_S$. The corresponding position in $\LCP_R$ is $p_i = W_p[i] = 5$
(or rather $p_i = s_i + W_r[i] = 5$), and the mapped position $j$ is
$j' = p_i+j-s_i = 7$. Finally, $\LCP_S[s_i-1] = W_c[i-1]=1$.
According to our formula, then, we have
$\LCP_S[8] = \LCP_S[s_i-1] + \LCP_R[j'] - \LCP_R[p_i-1] = 1+2-1=2$.

\subsection{Supporting \nsv/\psv/\rmq{} queries}

Suffix tree topology can be inferred from the \LCP{} array with range minimum
queries (\rmq) and next/previous smaller value (\nsv/\psv) queries
\cite{Fischer2009a}. Some suffix tree operations are more efficient
if we also support \emph{next/previous smaller or equal value} (\nsev/\psev)
queries \cite{Abeliuk2013}. Query $\mnsev(i)$ ($\mpsev(i)$) finds the next
(previous) value smaller than or equal to $\mLCP[i]$.

In order to support the queries, we build a $64$-ary \emph{minima tree} over
the phrases of the \RLZ{} parsing. Each leaf node stores the smallest \LCP{}
value in the corresponding phrase, while each internal node stores the
smallest value in the subtree. Internal nodes are created and stored in a
levelwise fashion, so that each internal node, except perhaps the rightmost
one of each level, has $64$ children.

We encode the minima tree as two arrays. The smallest \LCP{} values are
stored in $M_{\mLCP}$, which we encode as an \slarray. Plain array $M_{L}$
stores the starting offset of each level in $M_{\mLCP}$, with the leaves
stored starting from offset $M_{L}[1] = 1$. If $i$ is a minima tree node
located at level $j$, the corresponding minimum value is $M_{\mLCP}[i]$, the
parent of the node is $M_{L}[j+1] + \lfloor (i - M_{L}[j]) / 64 \rfloor$,
and its first child is $M_{L}[j-1] + 64 \cdot (i - M_{L}[j])$.

A range minimum query $\mrmq(sp,ep)$ starts by finding the minimal range of
phrases $w_{l}, \dotsc, w_{r}$ covering the query and the maximal range of
phrases $w_{l'}, \dotsc, w_{r'}$ contained in the query (note that $l \le l' \le
l+1$ and $r-1 \le r' \le r$). We then use the
minima tree to find the leftmost minimum value $j = M_{\mLCP}[k]$ in
$M_{\mLCP}[l',r']$, and find the leftmost occurrence $\mLCP[i] = j$ in phrase
$w_{k}$. If $l < l'$ and $M_{\mLCP}[l] \le j$, we decompress phrase $w_{l}$
and find the leftmost minimum value $\mLCP[i'] = j'$ (with $i' \ge sp$) in the
phrase. If $j' \le j$, we update $(i,j) \leftarrow (i',j')$. Finally we check
phrase $w_{r}$ in a similar way, if $r > r'$ and $M_{\mLCP}[r] < j$. The answer
to the range minimum query is $\mLCP[i] = j$, so we return
$(i,j)$.\footnote{The definition of the query only calls for the leftmost
minimum position $i$. We also return $\mLCP[i] = j$, because suffix tree
operations often need it.} Finally, the particular case where no phrase is
contained in $[sp,ep]$ is handled by sequentially scanning one or two phrases
in $\LCP$.

The remaining queries are all similar to each other. In order to answer query
$\mnsv(i)$, we start by finding the phrase $w_{k}$ containing position $i$,
and then determining $\mLCP[i]$. Next we scan the rest of the phrase to see
whether there is a smaller value $\mLCP[j] < \mLCP[i]$ later in the phrase. If
so, we return $(j,\mLCP[j])$. Otherwise we traverse the minima tree to find
the smallest $k' > k$ with $M_{\mLCP}[k'] < \mLCP[i]$. We decompress
phrase $w_{k'}$, find the leftmost position $j$ with $\mLCP[j] < \mLCP[i]$,
and return $(j,\mLCP[j])$.

%% file: exper.tex
\section{Experiments}

We have implemented the relative suffix tree in C++, extending the
old relative FM-index implementation.\footnote{The current implementation is available
at \url{https://github.com/jltsiren/relative-fm}.} The implementation is based
on the \emph{Succinct Data Structure Library (SDSL) 2.0}~\cite{Gog2014b}. Some
parts of the implementation have been parallelized using \emph{OpenMP} and the
\emph{libstdc++ parallel mode}.

As our reference \CSA{}, we used the \emph{succinct suffix array} (\SSA{})
\cite{Ferragina2007a,Maekinen2005} implemented using SDSL components. Our
implementation is very similar to \texttt{csa\_wt} in SDSL, but we needed
better access to the internals than what the SDSL interface
provides. \SSA{} encodes the Burrows-Wheeler transform as a \emph{Huffman-shaped
wavelet tree}, combining fast queries with size close to the
\emph{order\nobreakdash-$0$ empirical entropy}. This makes it the
index of choice for DNA sequences \cite{Ferragina2009a}. In addition to
the plain \SSA{} with uncompressed bitvectors, we also used \SSArrr{} with
entropy-compressed bitvectors \cite{Raman2007} to highlight the
the time-space trade-offs achieved with better compression

We sampled \SA{} in suffix order and \ISA{} in text order. In \SSA, the sample
intervals were $17$ for \SA{} and $64$ for \ISA. In \RFM, we used sample
interval $257$ for \SA{} and $512$ for \ISA{} to handle the regions that do
not exist in the reference. The sample intervals for suffix order sampling
were primes due to the long runs of character $N$ in the assembled genomes. If
the number of long runs of character $N$ in the indexed sequence is even, the
lexicographic ranks of almost all suffixes in half of the runs are odd, and
those runs are almost completely unsampled. This can be avoided by making the
sample interval and the number of runs \emph{relatively prime}.

The experiments were done on a system with two 16\nobreakdash-core AMD Opteron 6378 processors and 256~GB of memory. The system was running Ubuntu 12.04 with Linux kernel 3.2.0. We compiled all code with g++ version~4.9.2. We allowed index construction to use multiple threads, while confining the query benchmarks to a single thread. As AMD Opteron uses a \emph{non-uniform memory access} architecture, accessing local memory controlled by the same physical CPU is faster than accessing remote memory controlled by another CPU. In order to ensure that all data structures are in local memory, we set the CPU affinity of the query benchmarks with the \texttt{taskset} utility.

As our target sequence, we used the \emph{maternal haplotypes} of the
\emph{1000 Genomes Project individual NA12878}~\cite{Rozowsky2011}. As the
reference sequence, we used the 1000 Genomes Project version of the \emph{GRCh37
assembly} of the \emph{human reference
genome}.\footnote{\url{ftp://ftp.1000genomes.ebi.ac.uk/vol1/ftp/technical/reference/}}
Because NA12878 is female, we also created a reference sequence without
chromosome~Y.

In the following, a basic FM-index is an index supporting only \find{}
queries, while a full index also supports \locate{} and \extract{} queries.

\subsection{Indexes and their sizes}

Table~\ref{table:construction} lists the resource requirements for building
the relative indexes, assuming that we have already built the corresponding
non-relative structures for the sequences. As a comparison, building an
FM-index for a human genome typically takes 16--17 minutes and 25--26~GB
of memory. While the construction of the basic \RFM{} index is
highly optimized, the other construction algorithms are just the first
implementations. Building the optional \rselect{} structures takes 4 minutes
using two threads and around 730~megabytes ($\abs{R} + \abs{S}$ bits) of
working space in addition to \RFM{} and \rselect.

\begin{table*}
\caption{Sequence lengths and resources used by index construction for NA12878
relative to the human reference genome with and without chromosome~Y. Approx
and Inv denote the approximate \LCS{} and the bwt-invariant subsequence.
Sequence lengths are in millions of base pairs, while construction resources
are in minutes of wall clock time and gigabytes of
memory.}\label{table:construction}
\setlength{\extrarowheight}{2pt}
\setlength{\tabcolsep}{3pt}
\begin{center}
\begin{tabular}{c|cccc|cc|cc|cc}
\hline
 &
\multicolumn{4}{c|}{\textbf{Sequence length}} &
\multicolumn{2}{c|}{\textbf{\RFM{} (basic)}} &
\multicolumn{2}{c|}{\textbf{\RFM{} (full)}} &
\multicolumn{2}{c}{\textbf{\RCST}} \\
\textbf{ChrY} &
\textbf{Reference} & \textbf{Target} & \textbf{Approx} & \textbf{Inv} &
\textbf{Time} & \textbf{Memory} &
\textbf{Time} & \textbf{Memory} &
\textbf{Time} & \textbf{Memory} \\
\hline
yes & 3096M & 3036M & 2992M & 2980M & 1.42 min & 4.41 GB & 175 min & 84.0 GB &
629 min & 141 GB \\
no  & 3036M & 3036M & 2991M & 2980M & 1.33 min & 4.38 GB & 173 min & 82.6 GB &
593 min & 142 GB \\
\hline
\end{tabular}
\end{center}
\end{table*}

The sizes of the final indexes are listed in Table~\ref{table:indexes}. The
full \RFM{} is over twice the size of the basic index, but still
3.3\nobreakdash––3.7~times smaller than the full \SSArrr{} and
4.6\nobreakdash––5.3~times smaller than the full \SSA. The \RLCP{} array is
2.7~times larger than the \RFM{} index with the full human reference and
1.5~times larger with the female reference. Hence having a separate female
reference is worthwhile, if there are more than a few female genomes among
the target sequences. The optional \rselect{} structure is almost as large
as the basic \RFM{} index.

\begin{table*}
\caption{Various indexes for NA12878 relative to the human reference genome
with and without chromosome~Y. The total for \RCST{} includes the full \RFM.
Index sizes are in megabytes and in bits per character.}\label{table:indexes}
\setlength{\extrarowheight}{2pt}
\setlength{\tabcolsep}{3pt}
\begin{center}
\begin{tabular}{c|cc|cc|cc|ccc}
\hline
 &
\multicolumn{2}{c|}{\textbf{\SSA}} &
\multicolumn{2}{c|}{\textbf{\SSArrr}} &
\multicolumn{2}{c|}{\textbf{\RFM}} &
\multicolumn{3}{c}{\textbf{\RCST}} \\
\textbf{ChrY} &
\textbf{Basic} & \textbf{Full} &
\textbf{Basic} & \textbf{Full} &
\textbf{Basic} & \textbf{Full} &
\textbf{\RLCP} & \textbf{Total} & \textbf{\rselect} \\
\hline
\multirow{2}{*}{yes}
&  1248 MB &  2110 MB &   636 MB &  1498 MB &   225 MB &   456 MB &  1233 MB &  1689 MB &   190 MB \\
& 3.45 bpc & 5.83 bpc & 1.76 bpc & 4.14 bpc & 0.62 bpc & 1.26 bpc & 3.41 bpc & 4.67 bpc & 0.52 bpc \\
\hline
\multirow{2}{*}{no}
&  1248 MB &  2110 MB &   636 MB &  1498 MB &   186 MB &   400 MB &   597 MB &   997 MB &   163 MB \\
& 3.45 bpc & 5.83 bpc & 1.76 bpc & 4.14 bpc & 0.51 bpc & 1.11 bpc & 1.65 bpc & 2.75 bpc & 0.45 bpc \\
\hline
\end{tabular}
\end{center}
\end{table*}

Table~\ref{table:rfm components} lists
the sizes of the individual components of the relative FM-index.
Including the chromosome~Y in the reference increases the sizes
of almost all relative components, with the exception of $\mCS(\mBWT_{S})$ and
$\mAli(R,S)$. In the first case, the common subsequence still covers
approximately the same positions in $\mBWT_{S}$ as before. In the second case,
chromosome~Y appears in bitvector $B_{R}$ as a long run of \zerobit{}s, which
compresses well. The components of a full \RFM{} index are larger
than the corresponding components of a basic \RFM{} index, because the
bwt-invariant subsequence is shorter than the approximate longest
common subsequence (see Table~\ref{table:construction}).

\begin{table*}
\caption{Breakdown of component sizes in the \RFM{} index for NA12878 relative
to the human reference genome with and without chromosome~Y in bits per
character.}\label{table:rfm components}
\setlength{\extrarowheight}{2pt}
\setlength{\tabcolsep}{3pt}
\begin{center}
\begin{tabular}{c|cc|cc}
\hline
 & \multicolumn{2}{c|}{\textbf{Basic \RFM}} & \multicolumn{2}{c}{\textbf{Full
\RFM}} \\
\textbf{ChrY}                &      \textbf{yes} &       \textbf{no} &      \textbf{yes} &       \textbf{no} \\
\hline
\textbf{\RFM}                & \textbf{0.62 bpc} & \textbf{0.51 bpc} & \textbf{1.26 bpc} & \textbf{1.11 bpc} \\
$\mCS(\mBWT_{R})$            &          0.12 bpc &          0.05 bpc &          0.14 bpc &          0.06 bpc \\
$\mCS(\mBWT_{S})$            &          0.05 bpc &          0.05 bpc &          0.06 bpc &          0.06 bpc \\
$\mAli(\mBWT_{R},\mBWT_{S})$ &          0.45 bpc &          0.42 bpc &          0.52 bpc &          0.45 bpc \\
$\mAli(R,S)$                 &                -- &                -- &          0.35 bpc &          0.35 bpc \\
\SA{} samples                &                -- &                -- &          0.12 bpc &          0.12 bpc \\
\ISA{} samples               &                -- &                -- &          0.06 bpc &          0.06 bpc \\
\hline
\end{tabular}
\end{center}
\end{table*}

The size breakdown of the \RLCP{} array can be seen in Table~\ref{table:rlcp components}.
Phrase pointers and phrase lengths take space proportional to the number of phrases. As
there are more mismatches between the copied substrings with the full human reference
than with the female reference, the absolute \LCP{} values take a larger proportion of the
total space with the full reference. Shorter phrase length increases the likelihood that
the minimal \LCP{} value in a phrase is a large value, increasing the size of the minima tree.

\begin{table*}
\caption{Breakdown of component sizes in the \RLCP{} array for NA12878 relative
to the human reference genome with and without chromosome~Y. The number of phrases,
average phrase length, and the component sizes in bits per character. ``Parse''
contains $W_{r}$ and $W_{\ell}$, ``Literals'' contains $W_{c}$ and $L$, and ``Tree''
contains $M_{\mLCP}$ and $M_{L}$.}\label{table:rlcp components}
\setlength{\extrarowheight}{2pt}
\setlength{\tabcolsep}{3pt}
\begin{center}
\begin{tabular}{c|cc|ccc|c}
\hline
\textbf{ChrY} & \textbf{Phrases} & \textbf{Length} & \textbf{Parse} & \textbf{Literals} & \textbf{Tree} & \textbf{Total} \\
\hline
yes           &      128 million &            23.6 &      1.35 bpc &          1.54 bpc &      0.52 bpc &        3.41 bpc \\
no            &       94 million &            32.3 &      0.97 bpc &          0.41 bpc &      0.27 bpc &        1.65 bpc \\
\hline
\end{tabular}
\end{center}
\end{table*}

In order to use relative data structures, we also need to have the reference
data structures in memory. The basic \SSA{} used by the basic
\RFM{} takes 1283~MB with chromosome~Y and 1248~MB without, while the full
\SSA{} used by the full \RFM{} takes 2162~MB and 2110~MB, respectively. The
reference \LCP{} array used by the \RLCP{} array requires 3862~MB and 3690~MB
with and without chromosome~Y.

\subsection{Query times}

Average query times for the basic operations can be seen in Tables~\ref{table:rfm
queries} and~\ref{table:rlcp queries}. The results for \LF{} and \Psiop{} queries
in the full FM-indexes are similar to the earlier ones with basic indexes
\cite{Boucher2015}. Random access to the \RLCP{} array is about 30~times
slower than to the \LCP{} array, while sequential access is 10~times slower.
The \nsv, \psv, and \rmq{} queries are comparable to 1\nobreakdash--2~random
accesses to the \RLCP{} array.

\begin{table*}
\caption{Average query times in microseconds for 10~million random queries in the full
\SSA, the full \SSArrr, and the full \RFM{} for NA12878 relative to
the human reference genome with and without chromosome~Y.}\label{table:rfm
queries}
\setlength{\extrarowheight}{2pt}
\setlength{\tabcolsep}{3pt}
\begin{center}
\begin{tabular}{c|cc|cc|cc|c}
\hline
& \multicolumn{2}{c|}{\textbf{\SSA}} & \multicolumn{2}{c|}{\textbf{\SSArrr}} & \multicolumn{2}{c|}{\textbf{\RFM}} & \textbf{\rselect} \\
\textbf{ChrY} & \textbf{\LF} & \textbf{\Psiop} & \textbf{\LF} & \textbf{\Psiop} & \textbf{\LF} & \textbf{\Psiop} & \textbf{\Psiop} \\
\hline
yes           &   0.328 \mus &      1.048 \mus &   1.989 \mus &      2.709 \mus &   3.054 \mus &     43.095 \mus &      5.196 \mus \\
no            &   0.327 \mus &      1.047 \mus &   1.988 \mus &      2.707 \mus &   2.894 \mus &     40.478 \mus &      5.001 \mus \\
\hline
\end{tabular}
\end{center}
\end{table*}

\begin{table*}
\caption{Query times in microseconds in the \LCP{} array (\slarray) and the
\RLCP{} array for NA12878 relative to the human reference genome with and without
chromosome~Y. For the random queries, the query times are averages over
100~million queries. The range lengths for the \rmq{} queries were $16^{k}$ (for
$k \ge 1$) with probability $0.5^{k}$. For sequential access, we list the average
time per position for scanning the entire array.}\label{table:rlcp
queries}
\setlength{\extrarowheight}{2pt}
\setlength{\tabcolsep}{3pt}
\begin{center}
\begin{tabular}{c|cc|ccccc}
\hline
& \multicolumn{2}{c|}{\textbf{\LCP{} array}} & \multicolumn{5}{c}{\textbf{\RLCP{} array}} \\
\textbf{ChrY} & \textbf{Random} & \textbf{Sequential} & \textbf{Random} & \textbf{Sequential} & \textbf{\nsv} & \textbf{\psv} & \textbf{\rmq} \\
\hline
yes           &      0.054 \mus &          0.002 \mus &      1.580 \mus &          0.024 \mus &    1.909 \mus &    1.899 \mus &    2.985 \mus \\
no            &      0.054 \mus &          0.002 \mus &      1.480 \mus &          0.017 \mus &    1.834 \mus &    1.788 \mus &    3.078 \mus \\
\hline
\end{tabular}
\end{center}
\end{table*}

\begin{figure*}
\begin{center}
\includegraphics{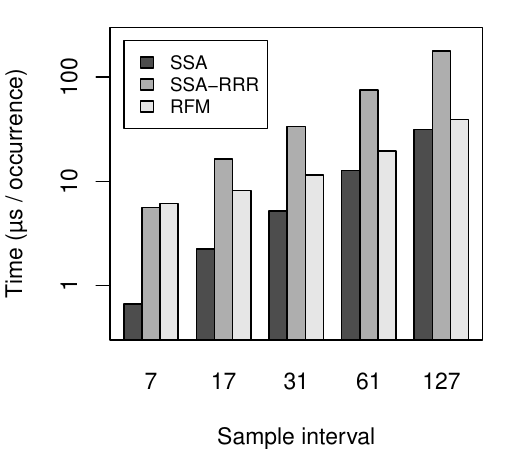}%
\hspace{-0.6in}%
\includegraphics{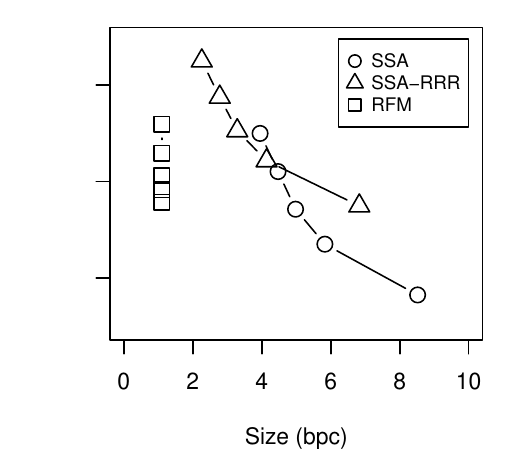}
\end{center}
\caption{Average \find{} and \locate{} times in microseconds per occurrence for 2~million patterns
of length $32$ with a total of 255~million occurrences on NA12878 relative to
the human reference genome without chromosome~Y. Left: Query time vs.\ suffix array
sample interval. Right: Query time vs.\ index size in bits per character.}\label{fig:locate}
\end{figure*}

We also tested the \locate{} performance of the full \RFM{} index, and
compared it to \SSA{} and \SSArrr. We built the indexes with \SA{} sample
intervals $7$, $17$, $31$, $61$, and $127$, using the reference without
chromosome~Y for \RFM.\footnote{With \RFM, the sample intervals apply
to the reference \SSA.} The \ISA{} sample interval was the
maximum of $64$ and the \SA{} sample interval. We extracted 2~million
random patterns of length $32$, consisting of characters $ACGT$, from
the target sequence, and measured the total time taken by \find{} and
\locate{} queries. The results can be seen in
Figure~\ref{fig:locate}. While \SSA{} and \SSArrr{} query times were
proportional to the sample interval, \RFM{} used 5.4\nobreakdash--7.6~microseconds
per occurrence more than \SSA{}, resulting in slower growth in query times.
In particular, \RFM{} with sample interval $31$ was faster than
\SSA{} with sample interval $61$. As the \locate{} performance of the \RFM{}
index is based on the sample interval in the reference, it is generally best
to use dense sampling (e.g.~7 or 17), unless there are only a few target
sequences.

\subsection{Synthetic collections}

In order to determine how the differences between the reference sequence and
the target sequence affect the size of relative structures, we built \RCST{}
for various \emph{synthetic datasets}. We took a 20~MB prefix of the human
reference genome as the reference sequence, and generated 25 target
sequences with every \emph{mutation rate} $p \in \set{0.0001, 0.0003, 0.001, 0.003,
0.01, 0.03, 0.1}$. A total of 90\% of the mutations were single-character
substitutions, while 5\% were insertions and another 5\% deletions. The length of an
insertion or deletion was $k \ge 1$ with probability $0.2 \cdot 0.8^{k-1}$.

\begin{figure*}
\begin{center}
\includegraphics{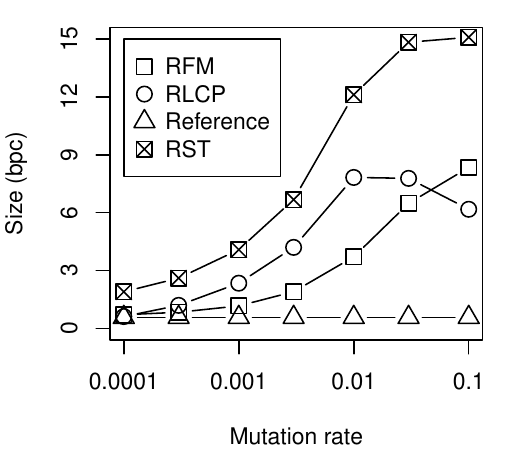}%
\hspace{-0.6in}%
\includegraphics{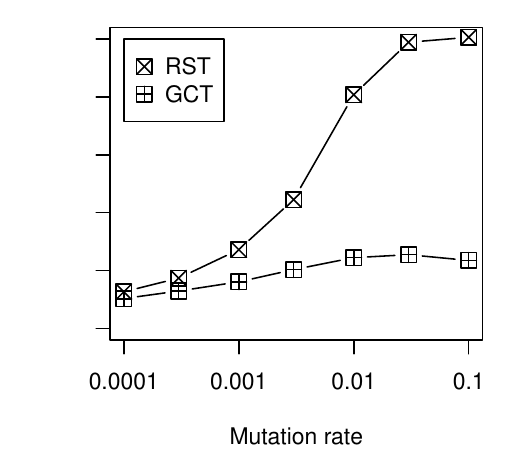}
\end{center}
\caption{Index size in bits per character vs.\ mutation rate for 25
synthetic sequences relative to a 20~MB reference.}\label{fig:synthetic}
\end{figure*}

The results can be seen in Figure~\ref{fig:synthetic}~(left). The size of the \RLCP{}
array grew quickly with increasing mutation rates, peaking at $p = 0.01$.
At that point, the average length of an \RLZ{} phrase was comparable to what
could be found in the \DLCP{} arrays of unrelated DNA sequences. With even
higher mutation rates, the phrases became slightly longer due to the smaller
average \LCP{} values. The \RFM{} index, on the other hand, remained small
until $p = 0.003$. Afterwards, the index started growing quickly, eventually
overtaking the \RLCP{} array.

We also compared the size of the relative suffix tree to \GCT{} \cite{Navarro2015},
which is essentially a \CSTsada{} for repetitive collections.
While the structures are intended for different purposes, the comparison shows how
much additional space is used for providing access to the suffix trees
of individual datasets. We chose to skip the \CSTnpr{} for repetitive collections
\cite{Abeliuk2013}, as its implementation was not stable enough.

Figure~\ref{fig:synthetic}~(right) shows the sizes of the compressed suffix
trees. The numbers for \RCST{} include individual indexes for each of the 25
target sequences as well as the reference data, while the numbers for \GCT{}
are for a single index containing the 25 sequences. With low mutation rates,
\RCST{} was not much larger than \GCT{}. The size of \RCST{} starts growing
quickly at around $p = 0.001$, while the size of \GCT{} stabilizes at
3\nobreakdash--4~bpc.

\subsection{Suffix tree operations}

In the final set of experiments, we compared the performance of \RCST{} to the
SDSL implementations of various compressed suffix trees. We used the maternal
haplotypes of NA12878 as the target sequence and the human reference genome
without chromosome~Y as the reference sequence. We built \RCST, \CSTsada,
\CSTnpr, and \FCST{} for the target sequence. \CSTsada{} uses \emph{Sadakane's
compressed suffix array} (\CSAsada) \cite{Sadakane2003} as its \CSA, while the
other SDSL implementations use \SSA. We used \PLCP{} as the \LCP{} encoding
with both \CSTsada{} and \CSTnpr{}, and also built \CSTnpr{} with \LCPdac.

We used three algorithms for the performance comparison. The first algorithm is
\emph{preorder traversal} of the suffix tree using SDSL iterators
(\texttt{cst\_dfs\_const\_forward\_iterator}). The iterators use operations
$\mRoot$, $\mLeaf$, $\mParent$, $\mFChild$, and $\mNSibling$, though $\mParent$
queries are rare, as the iterators cache the most recent parent nodes.

The other two algorithms find the \emph{maximal substrings} of the query string
occurring in the indexed text, and report the lexicographic range for each such
substring. This is a key task in common problems such as computing
\emph{matching statistics} \cite{Chang1994} or finding \emph{maximal exact matches}.
The \emph{forward algorithm} uses $\mRoot$, $\mSDepth$, $\mSLink$, $\mChild$,
and $\mLetter$, while the \emph{backward algorithm} \cite{Ohlebusch2010a} uses
$\mLF$, $\mParent$, and $\mSDepth$.

We used the \emph{paternal haplotypes} of chromosome~1 of NA12878 as the
query string in the maximal substrings algorithms. Because some tree operations
in the SDSL compressed suffix trees take time proportional to the depth of the
current node, we truncated the runs of character $N$ in the query string into
a single character. Otherwise searching in the deep subtrees would have
made some SDSL suffix trees much slower than \RCST.

The results can be seen in Table~\ref{table:cst}. \RCST{} was 1.8~times smaller
than \FCST{} and several times smaller than the other compressed suffix trees.
In depth-first traversal, \RCST{} was 4~times slower than \CSTnpr{} and
about 15~times slower than \CSTsada. \FCST{} was orders of magnitude slower,
managing to traverse only 5.3\% of the tree before the run was terminated after
24~hours.

\begin{table*}
\caption{Compressed suffix trees for the maternal haplotypes of NA12878
relative to the human reference genome without chromosome~Y. Component
choices; index size in bits per character; average time in microseconds per
node for preorder traversal; and average time in microseconds per character
for finding maximal substrings shared with the paternal haplotypes of
chromosome~1 of NA12878 using forward and backward algorithms. The figures in
parentheses are estimates based on the progress made in the first 24~hours.}\label{table:cst}
\setlength{\extrarowheight}{2pt}
\setlength{\tabcolsep}{3pt}
\begin{center}
\begin{tabular}{c|cc|c|c|cc}
\hline
& & & & & \multicolumn{2}{c}{\textbf{Maximal substrings}} \\
\textbf{\CST}      & \textbf{\CSA} & \textbf{\LCP} & \textbf{Size} & \textbf{Traversal} & \textbf{Forward} & \textbf{Backward} \\
\hline
\CSTsada           & \CSAsada      & \PLCP         &     12.33 bpc &          0.06 \mus &       79.97 \mus &         5.14 \mus \\
\CSTnpr            & \SSA          & \PLCP         &     10.79 bpc &          0.23 \mus &       44.55 \mus &         0.46 \mus \\
\CSTnpr            & \SSA          & \LCPdac       &     18.08 bpc &          0.23 \mus &       29.70 \mus &         0.40 \mus \\
\FCST              & \SSA          & --            &      4.98 bpc &      (317.30 \mus) &      332.80 \mus &         3.13 \mus \\
\RCST               & \RFM          & \RLCP         &      2.75 bpc &          0.90 \mus &      208.62 \mus &         3.72 \mus \\
\RCST{} + \rselect  & \RFM          & \RLCP         &      3.21 bpc &          0.90 \mus &       80.20 \mus &         3.71 \mus \\
\hline
\end{tabular}
\end{center}
\end{table*}

It should be noted that the memory access patterns of traversing \CSTsada, \CSTnpr, and \RCST{} are highly local. Traversal times are mostly based on the amount of computation done, while memory latency is less important than in the individual query benchmarks. In \RCST{}, the algorithm is essentially the following: 1) compute \rmq{} in the current range; 2) proceed recursively to the left subinterval; and 3) proceed to the right subinterval. This involves plenty of redundant work, as can be seen by comparing the traversal time (0.90~\mus{} per node) to sequential \RLCP{} access (0.017~\mus{} per position). A faster algorithm would decompress large parts of the \LCP{} array at once, build the corresponding subtrees in postorder \cite{Abouelhoda2004}, and traverse the resulting trees.

\RCST{} with \rselect{} is as fast as \CSTsada{} in the forward algorithm,
1.8\nobreakdash--2.7~times slower than \CSTnpr, and 4.1~times faster than \FCST. Without
the additional structure, \RCST{} becomes 2.6~times slower. As expected \cite{Ohlebusch2010a},
the backward algorithm is much faster than the forward algorithm. \CSTsada{} and \RCST,
which combine slow backward searching with a fast tree, have similar
performance to \FCST, which combines fast searching with a slow tree. \CSTnpr{} is about
an order of magnitude faster than the others in the backward algorithm.

%% file: concl.tex
\section{Discussion}\label{section:discussion}

We have introduced relative suffix trees (\RCST), a new kind of compressed suffix tree for repetitive sequence collections. Our \RCST{} compresses the suffix tree of an individual sequence relative to the suffix tree of a reference sequence. It combines an already known relative suffix array with a novel relative-compressed longest common prefix representation (\RLCP). When the sequences are similar enough (e.g., two human genomes), the \RCST{} requires about 3 bits per symbol on each target sequence. This is close to the space used by the most space-efficient compressed suffix trees designed to store repetitive collections in a single tree, but the \RCST{} provides a different functionality as it indexes each sequence individually. The \RCST{} supports query and navigation operations within a few microseconds, which is competitive with the largest and fastest compressed suffix trees.

The size of \RCST{} is proportional to the amount of sequence that is present either in the reference or in the target, but not both. This is unusual for relative compression, where any additional material in the reference is generally harmless. Sorting the suffixes in lexicographic tends to distribute the additional suffixes all over the suffix array, creating many mismatches between the suffix-based structures of the reference and the target. For example, the 60~million suffixes from chromosome~Y created 34~million new phrases in the RLZ parse of the \DLCP{} array of a female genome, doubling the size of the \RLCP{} array. Having multiple references (e.g.~male and female) can hence be worthwhile when building relative data structures for many target sequences.

While our \RCST{} implementation provides competitive time/space trade-offs, there is still much room for improvement. Most importantly, some of the construction algorithms require significant amounts of time and memory. In many places, we have chosen simple and fast implementation options, even though there could be alternatives that require significantly less space without being too much slower.

Our \RCST{} is a relative version of the \CSTnpr. Another alternative for future work is a relative \CSTsada, using \RLZ{} compressed bitvectors for suffix tree topology and \PLCP. 